\def\year{2018}\relax
\documentclass[letterpaper]{article} 
\usepackage{aaai18}  
\usepackage{times}  
\usepackage{helvet}  
\usepackage{courier}  
\usepackage{url}  
\usepackage{graphicx}  
\usepackage{algorithm}
\usepackage{algorithmic}
\usepackage{booktabs}
\usepackage{color}
\usepackage{ulem}

\definecolor{rd} {rgb} {1.0,0.0,0.0}
\definecolor{lg} {rgb} {0.7,0.3,0.9}
\definecolor{db} {rgb} {0.0,0.0,0.7}
\definecolor{dg} {rgb} {0.0,0.7,0.0}
\definecolor{dr} {rgb} {0.7,0.0,0.0}
\definecolor{gr} {rgb} {0.8,0.4,0.1}

\newcommand{\XY } [1] {{{\color{db}[XY] #1}}}
\newcommand{\YC   } [1] {{{\color{lg}[YC]#1}}}
\newcommand{\CH   } [1] {{{\color{dg}[CH]#1}}}
\newcommand{\final} {1}

\ifthenelse{\equal{\final}{1}} {\renewcommand{\XY }[1]{}}{}
\ifthenelse{\equal{\final}{1}} {\renewcommand{\YC }[1]{}}{}
\ifthenelse{\equal{\final}{1}} {\renewcommand{\CH }[1]{}}{}

\begin{document}
%
\title{Multiple Accounts Detection on Facebook \\Using Semi-Supervised Learning on Graphs}
\author{
Xiaoyun Wang, Chun-Ming Lai, Yunfeng Hong, Cho-Jui Hsieh, S. Felix Wu\\
University of California, Davis\\
Email: \{xiywang, cmlai, yfhong, chohsieh, sfwu\}@ucdavis.edu
}
\maketitle

    \section{Abstract}
    \label{sec:abstract}
    In social networks, a single user may create multiple accounts to spread his / her opinions and to influence others, by actively comment on different news pages. It would be beneficial to both social networks and their communities, to demote such abnormal activities, and the first step is to detect those accounts. However, the detection is challenging, because these accounts may have very realistic names and reasonable activity patterns. In this paper, we investigate three different approaches, and propose using graph embedding together with semi-supervised learning, to predict whether a pair of accounts are created by the same user. We carry out extensive experimental analyses to understand how changes in the input data and algorithmic parameters / optimization affect the prediction performance. We also discover that local information have higher importance than the global ones for such prediction, and point out the threshold leading to the best results. We test the proposed approach with 6700 Facebook pages from the Middle East, and achieve the averaged accuracy at 0.996 and AUC (area under curve) at 0.952 for users with the same name; with the U.S. 2016 election dataset, we obtain the best AUC at 0.877 for users with different names.
    

\section{Introduction}
\label{sec:introduction}
In the past decade, the number of people using online social networks (OSNs) as their sources of news and information has been growing rapidly. They not only receive information, but also share opinions. The cost of creating new accounts on OSNs is low, leading to the single-user multiple-accounts issue. People create multiple accounts for various reasons, and we focus on those using multiple accounts to comment on news pages. In that way, they build a false impression that their opinions are popular, in an attempt to influence others in the online communities \cite{King:2017:HTC}.

The multiple-accounts issue could also be raised by OSNs themselves. For instance, Facebook randomizes the account IDs when crawled by different crawler instances, as an anti-crawling feature.  \CH{not sure whether we need to mention this} We consider the OSN-introduced multiple accounts special cases of the multiple-accounts issue. \XY{I put this here just to explain why we have this issue. }

Multiple accounts are noises in datasets, and introduce inaccuracy in analyzed results, especially in user behavior studies. Detecting the multiple accounts cleans the data, and helps to improve the result quality of overall data analyses.

There are several existing approaches in multiple accounts detection, but they face some challenges:
\begin{itemize}
    \item Large portion of ground truth: previous works using supervised learning usually require a large portion of samples with ground truth, but getting ground truth is usually expensive. We can only query several hundreds accounts per day for the ground truth.
    \item Scalability: rich information, such as time stamps, IP addresses, and textual contents, may yield good detection results, but processing large amount of information imposes scalability issues.
    \item Exact number of users: GCN (graph convolutional network)~\cite{Kipf:2016:SSC}, previous semi-supervised works and clustering approaches \CH{cite those papers}  all need the exact number of users beforehand, which is usually unknown in practice.
\end{itemize}

In this paper, we investigate three methods to predict whether a pair of accounts belong to the same user:
\begin{itemize}
    \item unsupervised learning using Katz similarity;
    \item semi-supervised learning using Katz similarity;
    \item semi-supervised learning using graph embedding.
\end{itemize}
These methods use only a limited portion of ground truth, and who-comment-on-which-page information in the form of graphs, without knowing the exact number of users. 
To address the scalability issue, we also develop a clustering-based approach to reduce the search space, and use alternative ground truth to further reduce the number of ground truth queries.

We evaluate our methods with crawled pages from Facebook. With two small scale datasets, we compare the above three methods. We then extend to 100 datasets, each consists of accounts from the Middle East that shared the same display name. We also test the methods using news pages related to the 2016 United States election, with user activities randomly distributed into multiple split accounts, to simulate the effects of multiple accounts created by the same user. The obtained detection performance are reasonably good.

We make the following contributions : 
\begin{itemize}
    \item a novel semi-supervised method to detect multiple accounts in online social network, by using graph embedding to measure distances between nodes in graphs;  
    
    \item evaluations using large real-world datasets, with both OSN-introduced and user-introduced multiple accounts, showing state of the art performance;  
    
    \item experimental analyses to understand how the input data and algorithmic parameters and optimization affect the prediction performance.
\end{itemize}

\section{Related Work}
\label{sec:related}
Detecting multiple accounts in OSNs has been explored with user behavior analyses  and graph theory in recent years. Machine learning and artificial intelligence are also used for labeling nodes on social networks, such as labelling Wikipedia article categories. \cite{Tsikerdekis:2014:MAI}

User behavior analysis is an important tools for multiple-accounts detection. 
Tsikerdekis and Zeadally ~\shortcite{Tsikerdekis:2014:MAI} used nonverbal behaviors, such as time-dependent discussion of users, articles, and article discussions for multiple account identity deception detection. 
Gurajala et al. ~\shortcite{Gurajala:2016:PCF} utilized profile characteristics to detect fake Twitter accounts, based on account features, such as creation times, update times, the number of friends and followers.
Singh et al. ~\shortcite{Singh:2016:CPS} clustered users by their textual behaviors.
Sakakura et al. ~\shortcite{Sakakura:2012:DSB} measured similarity using bookmarks for spam detection.
User behavior analysis is based on profiling the basic features of users, and it inherently lacks deeper understanding in the interactions between users on OSNs. Moreover, it is possible to reverse engineer these methods, and generate statistical features that are similar to the normal accounts.

Graph based approaches can reveal the user-user, user-article, and user-page relationships on social networks.
Jiang et al. ~\shortcite{Jiang:2013:ULI} analyzed profile visit histories as well as links between passive profile and active comments, and built Latent interactions graphs to understand the behavior of users, as well as to detect fake accounts on OSNs.
Conti et al. ~\shortcite{Conti:2012:FDF} detected fake accounts using graph structures and longitudinal information.
Wang et al. ~\shortcite{Wang:2016:UCC} used clickstream and similarity graphs to cluster users, and then study users behaviors.
All these methods are based on graph clustering, which requires knowing how many users exist in the dataset. However, we do not know how many distinct users in our datasets. They can not tell whether two accounts belong to the same user. Only basic graph features, such as average degree, in/out edges, are used, but no deeper understanding of the graph is utilized. 

Machine learning models have also been applied in social network studies. \YC{In this paragraph, which methods are supervised, and which are semi-supervised?, the last one is semi, others are supervised learning}
Xiao and et al. ~\shortcite{Xiao:2015:DCF} developed an supervised learning approach to detect fake accounts registered by the same user, using IP addresses and registration dates provided by the LinkedIn dataset.
Logistic regression is widely used in prediction and classification.
For example, Zheng et al. ~\shortcite{Zheng:2015:DSS} used support vector machine (SVM) to detect spammers on social network. 
Naive bayes, decision trees, and random forest are popular approaches ~\cite{Fire:2012:SID,Lai:2017:AST,Boshmaf:2015:TFO} in detecting fake users and malicious users on social network too. 
Gong et al. ~\shortcite{Gong:2014:SSS} applied semi-supervised learning for sybil detection, by labeling sampled nodes as benign or sybil, then classified the nodes with information in the directed messages, together with known labels of the nodes. 
The supervised learning methods require a large number of ground truth to limit performance variances, and the semi-supervised methods for sybil detection requires label for sampled nodes. However, for our application, getting the ground truth is time consuming, and the nodes have no label. This makes those approaches not applicable. 

Graph embedding is an approach that could quantify nodes into vectors. It is widely used in link prediction, node classification, multi-label learning and clustering. Different graph embedding method captures different graph features.
Node2Vec proposed by Grover and Leskovec ~\shortcite{Grover:2016:NSF} is a method based on random walk; it provides a trade-off between global and local information. Choosing the right balance makes Node2Vec preserves community structures, as well as structural equivalences between nodes.  
Deep walk ~\cite{Perozzi:2014:DOL} introduced before Node2vec, which can considered as a special case with $p$ and $q$ (parameters in Node2vec) equal to 1.
SDNE ~\cite{DBLP:conf/kdd/WangC016} and DNGR ~\cite{DBLP:conf/aaai/CaoLX16} are auto-encoders which capture non-linearity in graphs.  
Cao et al. ~\shortcite{Cao:2015:GLG} used k-step probability matrix and matrix factorization to get global structure representations of graphs. \CH{this sentence is unclear to me} \XY{changed, to describe another graph embedding method called grarep}
All the deep learning methods above perform node classifications or multi-labeling, and the number of different classes or labels are known. However, in most real-world applications, such number is rarely available. 

\section{Data Description}
\label{sec:data}
The data we used are crawled from Facebook using the Social Interactive Networking and Conversation Entropy Engine (SINCERE) system ~\cite{Erlandsson:2015:COS} for more than five years, and consist of news pages from Asia, North America, the Middle East and Europe. The open-sourced SINCERE system uses distributed crawler instances to increase the crawling rate. However, as mentioned in the Introduction section, Facebook randomizes the account numbers for different crawler instances, leading to the OSN-introduced multiple accounts issue. These OSN-introduced cases are a little different from the user introduced ones: the OSN-introduced multiple accounts for the same user have the same display name, while those user-introduced ones may not. There are 10,044,228,650 accounts and 23,579,873 pages in the database. But there are only two billion users  on Facebook ~\cite{Constine:2017:NMA}, which means we are seeing about five accounts per user on averaged. 

We focus our study on the news pages \footnote{Our datasets are available at 
\url{anonymous_url}} 
from the Middle East and the 2016 US election. The Middle East data cover more than 6700 pages, and 100 datasets. Each dataset contains all the accounts sharing the same display name, together with all commenting activities by these accounts on the news pages. The sizes of these datasets range from 9k to 30k accounts, and 14k to 92k activities. We select data from the Middle East, because the users there have high tendency to share the same names, giving us good opportunities to test our methods. The 2016 US election dataset covers 34 news pages, with 6 million accounts and 26,985,976 commenting activities from these accounts. This dataset does not have OSN-introduced multi-accounts, because only one crawler instance is used to get the dataset.

\subsection{Ground Truth}
We write a separated crawler to query the unique user ID (the primary ID) of each account (identified by a scope ID) for the Middle East datasets. This crawler runs much slower than the SINCERE system, at only hundreds of accounts per day. Running the primary ID crawler for large number of accounts is not practical, but it tells us the ground truth about which accounts are indeed OSN-introduced multiples.


There is no good and guaranteed way to know which accounts are user-introduced multiples. Instead, we randomly separate activities of each account, which uniquely identified a single user in this particular dataset, into different split accounts, to simulate the user-introduced multiples. In this way, the ground truth is whether a pair of split accounts come from the same original account.

\section{Design}
\label{sec:design}
We introduce our approaches in this section: begin with how to contract relationship graphs from the crawled datasets, followed by three methods for multiple accounts detection, together with scalability analysis and optimization at the end of the section.

\subsection{Graph Construction}

We construct a graph $G(V, E)$ for each dataset, and carry out multiple accounts detection using the graph. The nodes $V$ represent the accounts and the news pages, and the edges $E$ represent relationships between accounts and pages: an edge $e$ exists when an account comments on or likes a news page. The constructed graphs are bipartite graphs, because edges only exist between accounts and pages, but not within accounts or pages themselves. We use undirected bipartite graphs, either weighted or unweighted, in our experiments. 
However, our approaches do not depend on these graph properties, and should be able to extend to more general graph types.

\begin{figure}
\centering
    \begin{tabular}{cc}
        \includegraphics[width=0.20\textwidth]{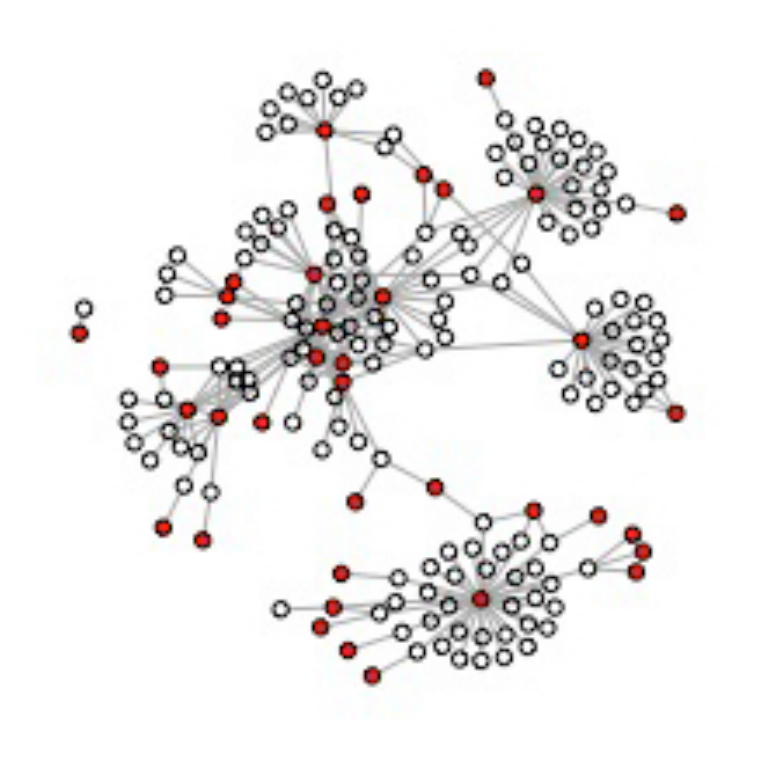}
    &
        \includegraphics[width=0.20\textwidth]{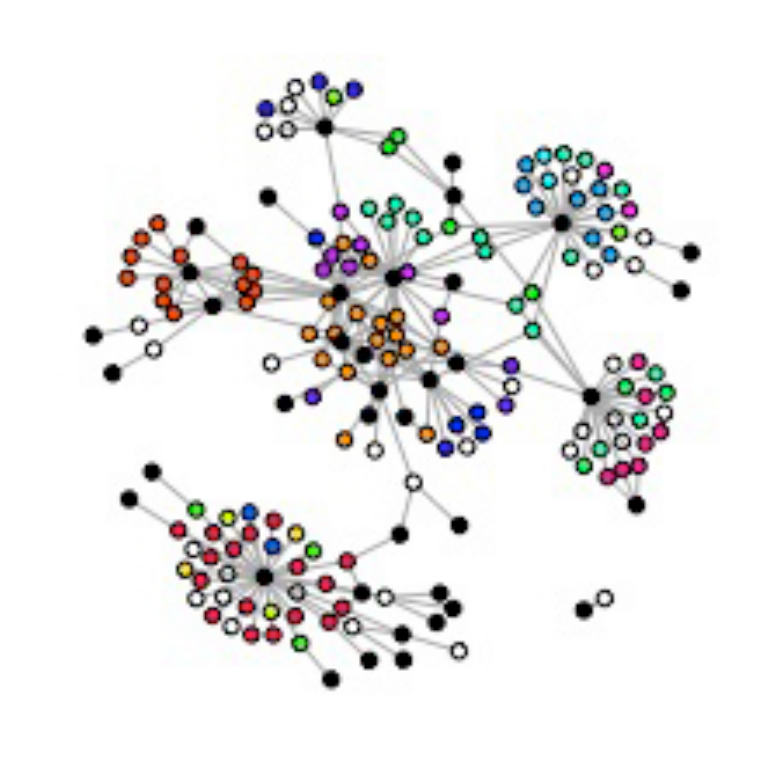}
    \end{tabular}
\caption{An example of a constructed graph. Left: the constructed graph, with accounts colored in white and pages colored in red; right: the detection results, with multiple accounts from the same user colored by the same color, except for black (pages) and white (users with only a single account each).
    \label{fig:graph}}
\end{figure}

Figure~\ref{fig:graph} gives an example based on a simple dataset, with the constructed graph on the left and the detection results on the right. A little bit of clustering can be seen from this example already, and we make use of the clustering results in our methods.

\subsection{Unsupervised Learning using Katz Similarity}

Unsupervised learning has the advantages of not requiring any ground truth, which is difficult to obtain. The main step of this method is to calculate similarities (i.e. distances) between different accounts. There are various matrices to measure the similarities, such as common neighbors, common edges, and node-edge scores. We choose Katz similarity ~\cite{Katz:1953:NSI} here. It is a commonly used topological measurement in social network studies, and Esfandar et al. ~\shortcite{Esfandar:2010:FKC} used it for linked prediction, thus we want to use it to predict whether an pair of accounts belongs to the same user. Katz similarity can be computed by:

\begin{equation}
(I-\beta M)^{-1} -I, 
\end{equation}
where $M$ is the adjacency matrix representation of $G$, $\beta$ is a scalar smaller than $1/\|M\|_2$ to ensure convergence, and $I$ is the identity matrix.

\begin{algorithm}
\caption{Unsupervised Method using Katz Similarity \label{algo:unsupervised} \CH{the algorithm box looks strange. check whether this is correct or not} \XY{yes you are right, should be percentile $\alpha$}}
    
    \begin{algorithmic}[1]
    \STATE compute the Katz similarity matrix $S$ of $G$;
    \STATE note the account nodes of $G$ as $V_a$;
    \IF {the similarity $S_{u, v}$ between two accounts $u$ and $v$ in $V_a$ is larger than an empirical threshold percentile $\alpha$ of $S$}
    \STATE $u$ and $v$ are predicted to be multiple accounts of the same user; 
    \ELSE 
    \STATE $u$ and $v$ are predicted to be accounts of different users.
    \ENDIF
    \end{algorithmic}
\end{algorithm}

The unsupervised method using Katz similarity is listed in Algorithm~\ref{algo:unsupervised}. It computes the similarity matrix to measure how closed two accounts are in the graph, and predicts they belong to the same user if the similarity is larger than a threshold percentile $\alpha$. $\alpha$ is a critical parameter in this method. It is selected empirically, and may need to change according to datasets. If a few ground truth are given, $\alpha$ can be selected by cross validation.
\CH{Maybe we can say $\alpha$ can be selected by cross validation if few ground truths are given?} \XY{OK}

\subsection{Semi-Supervised Learning using Katz Similarity}

In order to avoid selecting the values of $\alpha$, we turn to semi-supervised learning using the label spreading model ~\cite{zhou2004learning}. Unlike conventional label spreading methods which work on the labels of nodes, we predict the labels of node pairs. More precisely, we predict the value $L_{u,v}$ of pair $(u,v)$, which is defined as how likely accounts $u$ and $v$ belong to the same user; $L_{u,v}$ equals to $1$ if $u$ and $v$ are indeed multiple accounts of the same user, $0$ if they belong to different users, and $-1$ when unknown.

\begin{algorithm}
    \caption{Semi-Supervised Method using Katz Similarity \label{algo:supervised_katz}}
    \begin{algorithmic}[1]
        \STATE compute Katz similarity matrix $S$ of $G$;
        \STATE note the accounts nodes of $G$ as $V_a$;
        \STATE randomly sample $1/4$ of nodes in $V_a$ to query the ground truth;
        \STATE assign each $X_{u,v}$ with $max(S)-S_{u,v}+ \epsilon$;
        \STATE assign each $L_{u,v}$ with: 1 if accounts $u$ and $v$ are known to belong to the same user, 0 if known to belong to different users, and -1 if unknown;
        \STATE train the label spreading model with RBF kernel, using $X$ and $L$ as inputs;
        \STATE predict elements in $L$ which are initially unknown.
    \end{algorithmic}
\end{algorithm}

The semi-supervised method using Katz similarity is listed in Algorithm~\ref{algo:supervised_katz}. In the label spreading model, we use Radial basis function (RBF) kernel:  $K(x_1, x_2)= \exp (-\frac{ \|x_1 - x_2\|^2}{2\sigma^2})$, where $\sigma$ is the scaling parameter, and $x_i$ denotes the feature of a pair of nodes $(u, v)$. We use Katz similarity as a feature, so $x_i=S_{u,v}$ ($S$ is defined in $(I-\beta M)^{-1} -I$). \CH{I rewrote this. Check whether it is correct...} \XY{correct}
We  choose to sample $1/4$ of the nodes, aiming an 6.25\% label rate of input data. 

While this method does avoid knowing the values of $\alpha$, it yields poor prediction accuracy. We suspect only giving the label spreading model a single scalar value per edge, which is the Katz similarity, may not carry sufficient information to make good prediction by the model; thus, we try to feed more information into the model by using graph embedding. 

\subsection{Semi-Supervised Learning using Graph Embedding}

Graph embedding gives each node a vector, and we use Node2Vec ~\cite{Grover:2016:NSF} here. Node2Vec obtains the node embedding  based on random walks, and these walks can be more local or global by controlling a pair of parameters $p$ and $q$, where $p$ represent the likelihood to immediately revisit a node, and $q$ controls whether the random walks are more in breadth or depth directions. When $q > 1$, the random walks are more closed to the current nodes (i.e. walk locally); when $q < 1$, the random walks will visit nodes far from the current ones (i.e. walk globally).   \CH{can we explain what are p and q using one or two sentences?} \XY{added}  As a result, by choosing these two parameters our algorithm can also  balance the global or local information of graph. We will examine how these parameters affect the prediction accuracy in the upcoming Performance Analyses subsection. We also choose the size $d$ of each embedding vector to be 128, which is the default setting in Node2Vec.

\begin{algorithm}[tb]
    \caption{Semi-Supervised Method using Graph Embedding \label{algo:supervised_n2v}}
    \begin{algorithmic}[1]
        \STATE use Node2Vec to give each node $v$ of $G$ an embedding vector $W_v$;
        \STATE note the accounts nodes of $G$ as $V_a$;
        \STATE randomly sample $1/4$ of nodes in $V_a$ to query the ground truth;
        \STATE assign $X_{u,v}$ with $||W_u,W_v||_{\bar{1}}$;
        \STATE assign $L_{u,v}$ with: 1 if accounts $u$ and $v$ are known to belong to the same user, 0 if known to belong to different users, and -1 if unknown;
        \STATE train the label spreading model with RBF kernel, using $X$ and $L$ as input; 
        \STATE predict elements in $L$ which are initially unknown.
    \end{algorithmic}
\end{algorithm}

\begin{table}
    \centering
    \begin{tabular}{cc} 
        \toprule
        Operation & Definition\\
        \midrule
        Average & $[W_u \oplus W_v]_i = \frac{[W_u]_i + [W_v]_i}{2}$ \\ 
        Weighted L-1 norm & $[||W_u \cdot W_v||_{\bar{1}}]_i = |[W_u]_i - [W_v]_i|$ \\
        Weighted L-2 norm & $[||W_u \cdot W_v||_{\bar{2}}]_i = |[W_u]_i - [W_v]_i|^2$ \\
        Cosine & $W_u \cos W_v = \frac{W_u \cdot W_v}{||W_u|| ||W_v||}$ \\
        \bottomrule
    \end{tabular}
    \caption {Operations to measure similarities between two nodes with embedding vectors 
        \label{tab:operations}}
\end{table}

The semi-supervised method using graph embedding is listed in Algorithm~\ref{algo:supervised_n2v}. Compared to Algorithm~\ref{algo:supervised_katz} using Katz, the only difference is on how the similarities between nodes are measured, and subsequently how the distances between edges are calculated. Because each node now carries a length $d$ vector, there are several operations (listed in Table~\ref{tab:operations}) that can be used to measure the similarity between the embedding vectors $W_u$ and $W_v$ of two nodes $u$ and $v$. We try all the operations, and observe that L-1 norm and $\cos$ yield the best and similar accuracy, but L-2 norm and average do not perform so well in our tests.  

Graph embedding can extract more information from the graphs than Katz similarity, and brings better prediction accuracy. However these semi-supervised methods have drawbacks, mainly in terms of scalability. For datasets with thousands of accounts, getting ground truth on $1/4$ of them can be very time consuming. Moreover, the label spreading model runs on pairs of accounts. The number of pairs is usually millions or more since it grows quadratically with number of accounts, and this will result in high computational time and memory requirement. Thus, in the following section we will discuss how to scale up the proposed algorithms.  

\subsection{Scaling up the proposed algorithms}


We introduce a clustering-based approach to improve the scalability. The bottlenecks in our algorithm mainly exist at two places: 1) the computational time and memory requirement for label spreading,  and 2) the time-consuming ground truth queries. 
Both bottlenecks are related to the number of possible account pairs. Let $n$ be the number of accounts in $V_a$, thus there are $\frac{n^2}{2}$ pairs of accounts as inputs to the label spreading model.  \CH{do we need to say this here? Seems to be an implementation detail. } \XY{changed}.  The label spreading model runs in $O(n^6)$, with memory requirement in $M(n^4)$. If we can know with high certainty that some accounts are not from the same user, both bottlenecks can be overcome. The key idea is clustering: group accounts potentially belongs to the same users together, and separate those that do not. We use Spectral clustering ~\cite{Yu:2003:MSC} to cluster $V_a$ into sub-graphs based on the embedded vectors of nodes, then only query ground truth and use the label spreading model on account pairs within the same sub-graphs. For example if we use $c$ clusters, the computation complexity reduces to $O(\frac{n^6}{c^5})$, and the memory footprint reduces to $O(\frac{n^4}{c^3})$. In this way, we significantly reduce the workload of label spreading.

\begin{figure*}
    \centering
    \includegraphics[width=0.7\textwidth]{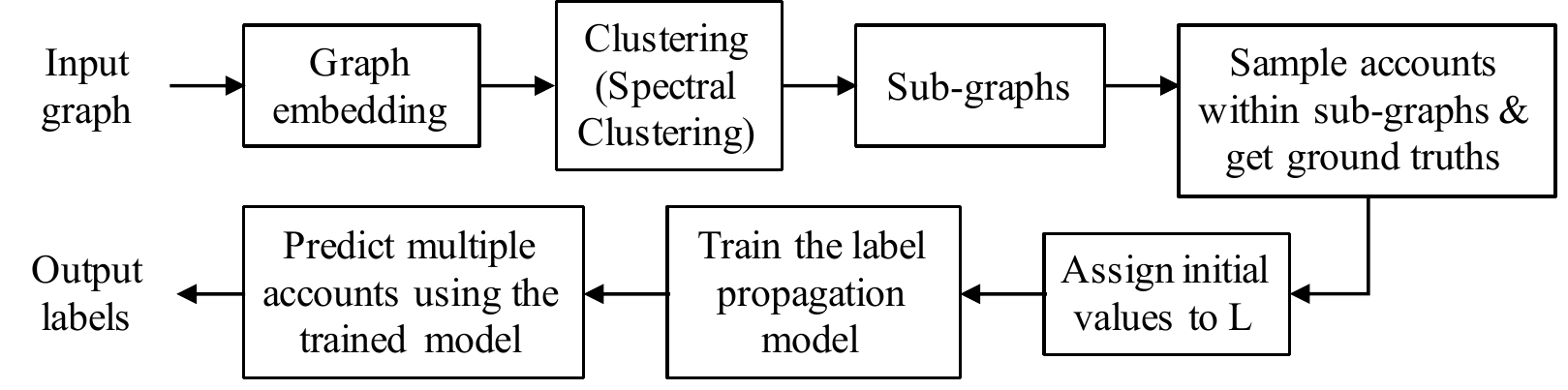}
    \caption{Flowchart showing all steps of Algorithm~\ref{algo:supervised_n2v} with optimizations.\label{fig:flowchat} \CH{maybe remove 1/4 in the flowchart} \XY{will remove}
    \YC{where is the alternative ground truths?}} \XY{get ground truth means get both "ground truth" and alternative ground truth, I use them together, Cho says do not need to say here}
\end{figure*}

We also use alternative ground truth to reduce the number of ground truth queries, and to improve scalability. Record from the Unsupervised Learning using Katz Similarity subsection, Katz similarity can roughly tell how closed two accounts are. By setting a high threshold (guaranteed to be higher than the empirical threshold), those pairs above such a threshold should belong to the same user, with high certainty; similarly, using a very low threshold could tell which pairs are very unlikely to come from the same user. 

Figure~\ref{fig:flowchat} shows the flow chart for our proposed semi-supervised learning method using graph embedding, with the graph clustering optimization.

\section{Experiments and Evaluation }
\label{sec:experiments}
In this section, we first compare the three methods introduced in the previous section with two small datasets, then apply semi-supervised learning using graph embedding on large datasets, and finally analysis how input datasets, algorithmic parameters, and optimization affect the prediction performance.

Because the prediction results are binary, we evaluate them using precision, recall, F1 score, accuracy and AUC (Area Under Curve, which measures the area under ROC curve, if the prediction is totally random, AUC would be 0.5). They are defined as following :
\begin{enumerate}
    \item $Precision = \frac{tp}{tp+fp}$
    \item $Recall = \frac{tp}{tp+fn}$
    \item $F_1 = \frac{2tp}{2tp+fp+fn}$ 
    \item $Accuracy = \frac{tp+fn}{tp+fp+tn+fn}$
    \item $AUC = \int TPR(T)(-FPR'(T)) dT$
\end{enumerate}
Here $tp$, $fp$, $tn$, $fn$ denote number of true positive, false positive, true negative and false negative, respectively. $TPR$ and $FPR$ represent true positive rate and false positive rate, which is defined as $\frac{tp}{tp+fn}$ and $\frac{fp}{fp+tn}$ respectively.

Data cleaning is necessary before running our approach. We want to filter out accounts with few actives, thus with little probability to come from the same user with other accounts. We project the bipartite graph $G$ between accounts and pages on the account side, using the rule: node $u$ and $v$ are connected with edge weight $w$ in projected graph $G_p$, if and only if  account $u$ and $v$ have $w$ common neighbors in graph $G$. After we get the projected graph $G_p$, we filter out the nodes with degree less or equal to $log_{10}|V|$. 

\subsection{Comparison among the Three Methods}

We use two simple datasets: one with 188 accounts and 262 activities, and the other with 4188 accounts and 6715 activities, to compare the three methods, and the results can be found in Table ~\ref{tab:method_comparsion}. Each dataset covers a unique display name from the Middle East pages, and they are not part of the large scale 100 names Middle East datasets. \CH{need to say what's the difference between two datasets. Different samples?} \XY{they are different names, the size are different, for Middle east data, there are 10K to 30K nodes in each graph. }

\begin{table}
    \centering
    \begin{tabular}{ccccc} 
        \toprule
        Dataset & Method & Precision & Recall & AUC\\ 
        \midrule
        1 & u. Katz & 0.58  & 0.81 & 0.84\\ 
        2 & u. Katz & 0.43& 0.78 & 0.86\\
        1 & s. Katz & 0.57  & 0.81 & 0.84\\ 
        2 & s. Katz &0.63  & 0.65  & 0.81 \\    
        1 & s. embedding & 0.77& 0.91 & 0.93 \\
        2 & s. embedding & 0.74& 0.74 & 0.86\\
        \bottomrule
    \end{tabular}
    \caption{Comparison among three methods. u. stands for unsupervised, and s. stands for semi-supervised. \label{tab:method_comparsion} \CH{no s.Katz for dataset 2?} \XY {added, and it not that bad }}
\end{table}

The performance of unsupervised method using Katz similarity is not bad. We use 75\% and 95\% as $\alpha$ for dataset 1 and 2, respectively. With more users in dataset 2, the response rate is lower than dataset 1 with fewer users, as expected. The main issue of this method is the requirement of setting the correct $\alpha$, and we will show how changes in $\alpha$ affect the prediction performance in the Performance Analyses subsection. In our experiments we find $\alpha$ using cross validation. However this method still has some advantages. It is faster than semi-supervised learning methods, because only matrix operations are used. It also helps to generate alternative ground truth, when the actual ground truth is not sufficient.

Semi-supervised learning method using Katz similarity matrix performs worse than semi-supervised learning method using graph embedding. We suspect the poor performance is caused by the lack of information feeding into the label spreading model, since only one scalar value is associated with each given account pair. 

Semi-supervised learning method using graph embedding brings promising results, with both datasets. Unlike the unsupervised method, this semi-supervised method utilize a small portion of ground truth. Furthermore, it feeds more information to the label spreading model, with a length-128 embedding vector  per account containing both local and global information, and a subsequent length-128 vector per account pair. Because of the good performance, we mainly focus on this method in the upcoming experiments. 

\subsection {Evaluation with Large Datasets}

We evaluate the semi-supervised learning using graph embedding method with two large scale datasets, one from the Middle East and the other from the 2016 U.S. election.

\begin{table}
    \centering
    \begin{tabular}{ccccc}
        \toprule
        precision & recall & F1 score & accuracy & AUC \\
        \midrule
        0.830 & 0.953 & 0.886 & 0.996 & 0.952   \\ 
        \bottomrule
    \end{tabular}
    \caption{Prediction performance for 100 display names in Middle East datasets \label{tab:100_names}}
\end{table}

\begin{figure*}
    \centering
    \includegraphics[width=1\textwidth]{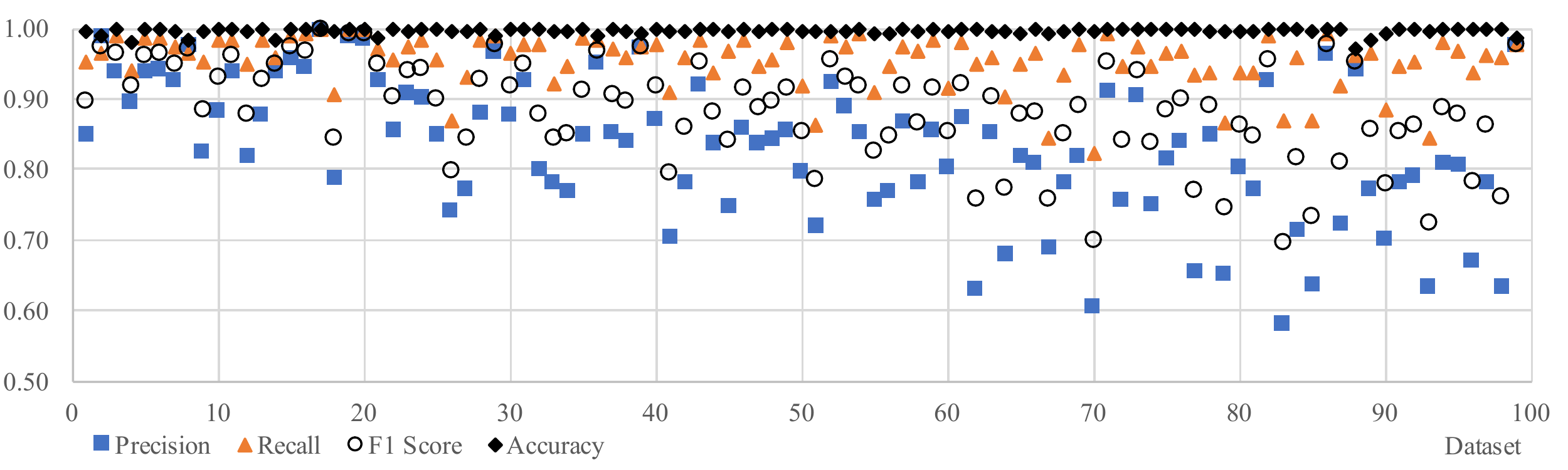}
    \caption{Prediction performance of 100 datasets from the Middle East, arranged in decreasing sizes, from left to right. 
   \label{fig:100names}}
\end{figure*}

The results with 100 datasets (i.e. display names) from the Middle East is summarized in Table ~\ref{tab:100_names}, and the individual results for each dataset are shown in Fig.~\ref{fig:100names}. We did not get the ground truth for all one million accounts, because that would take much longer time than practically possible; we used alternative ground truth instead \CH{what is alternative ground truths?}. \XY{alternative ground truth is that we use katz to generate some ground truth, because getting ground truth is very expensive} There are overlaps between the queried and the alternative ground truth, and we did not find conflicts in them. The overall performance is quite convincing, particularly with AUC at 0.952. When the number of accounts increases, the precision and the F1 score trend to  increase
, while the accuracy and recall are almost within a constant range.

\begin{table}
    \centering
    \begin{tabular}{c c c c c}
        \toprule
        precision & recall & F1 score & accuracy  & AUC \\
        \midrule
        0.780 & 0.756  & 0.768 &0.996 & 0.877 \\ 
        \bottomrule
    \end{tabular}
    \caption{Prediction performance of the U.S. 2016 dataset \label{tab:US_election}}
\end{table}


The results with the U.S. 2016 election dataset are listed in Table~\ref{tab:US_election}. We would like to detect active users using multiple accounts to comment on the news pages, users with less than 1000 comments are hence ignored. After distributing user activities into multiple split accounts (mentioned in the Ground Truth sub-section), our approach is able to achieve good prediction performance. This experiment is extremely challenging, because it is very difficult to differentiate two users who like to comment on similar set of pages, with only graph information. This is why the achieved AUC is slightly lower than other datasets contributed by the higher false negative rate, but we still have it at 0.88.

Our results show performance advantages over previous works. Although not running on the same datasets, we could still make some comparisons qualitatively. 
Cao et al. ~\shortcite{Xiao:2015:DCF} used supervised learning for fake account detection on LinkedIn datasets, and achieved 0.949 AUC, with all extra information such as IP addresses and account creation times. We achieve 0.952 AUC, with the Middle East datasets using only graph properties.
Tsikerdekis et al. ~\shortcite{Tsikerdekis:2014:MAI} used user revision activities in a specific time window, together with nonverbal information of articles and discussions, to detect multiple accounts in Wikipedia. Their precision, recall and F1 score are 0.729, 0.646 and 0.688 respectively. 
It can be seen that, although we only use the graph features, lacking other information used by prior works, our performance is comparable or even better. 


\subsection{Performance Analyses}

We carry out extensive experiments to further understand how the prediction performance changes, with regard to the properties of input datasets, as well as chosen parameter values in the algorithms. 

\subsubsection{Averaged Number of Activities per Account}
\begin{figure}
    \centering
    \includegraphics[width=0.47\textwidth]{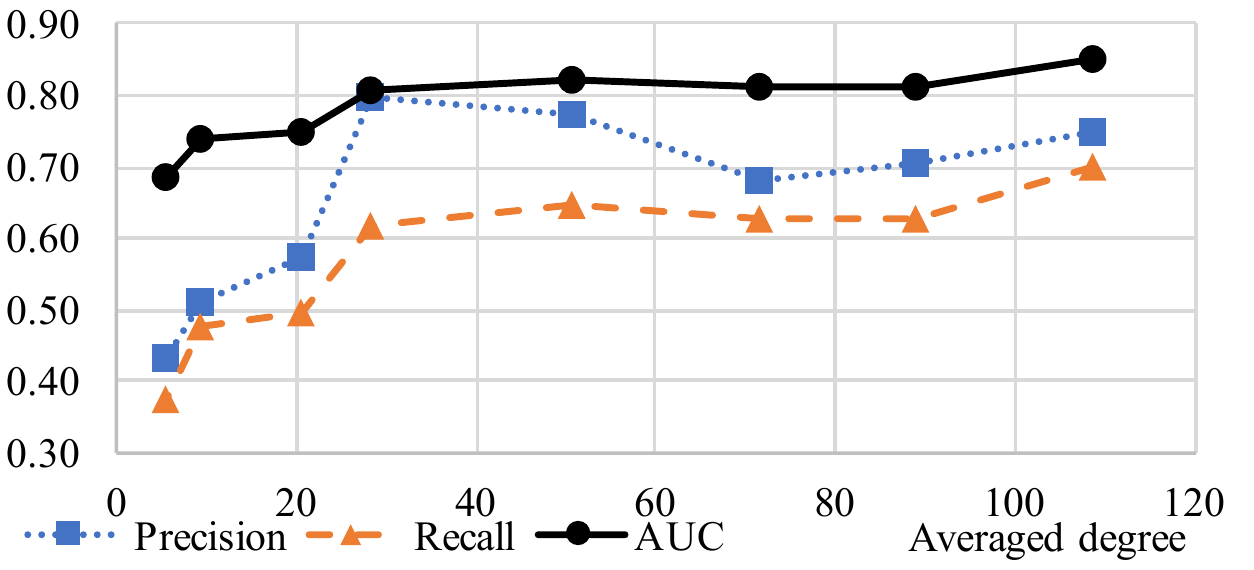}
    \caption{Prediction performance vs. averaged number of activities per account. \label{fig:density}}
\end{figure}

Figure~\ref{fig:density} shows how the averaged activity level affects the prediction performance. In this experiment, we sample certain level of activity per user from the U.S. 2016 dataset, then randomly assign the sampled activities of a user to 15 split accounts. The averaged number of activities per account is also the averaged degree of nodes in the constructed graph. It can be seen that generally when each account has more activities, i.e. when the constructed graph is more dense, the prediction performance is better. This is because more activities will lead to more informative node embedding vectors in our algorithm. However, the improvement in AUC is marginal when the averaged degree is larger than 30, and the best precision happens when the averaged degree is 30. 

\subsubsection{Averaged Number of Accounts per User}
\begin{figure}
    \centering
    \includegraphics[width=0.47\textwidth]{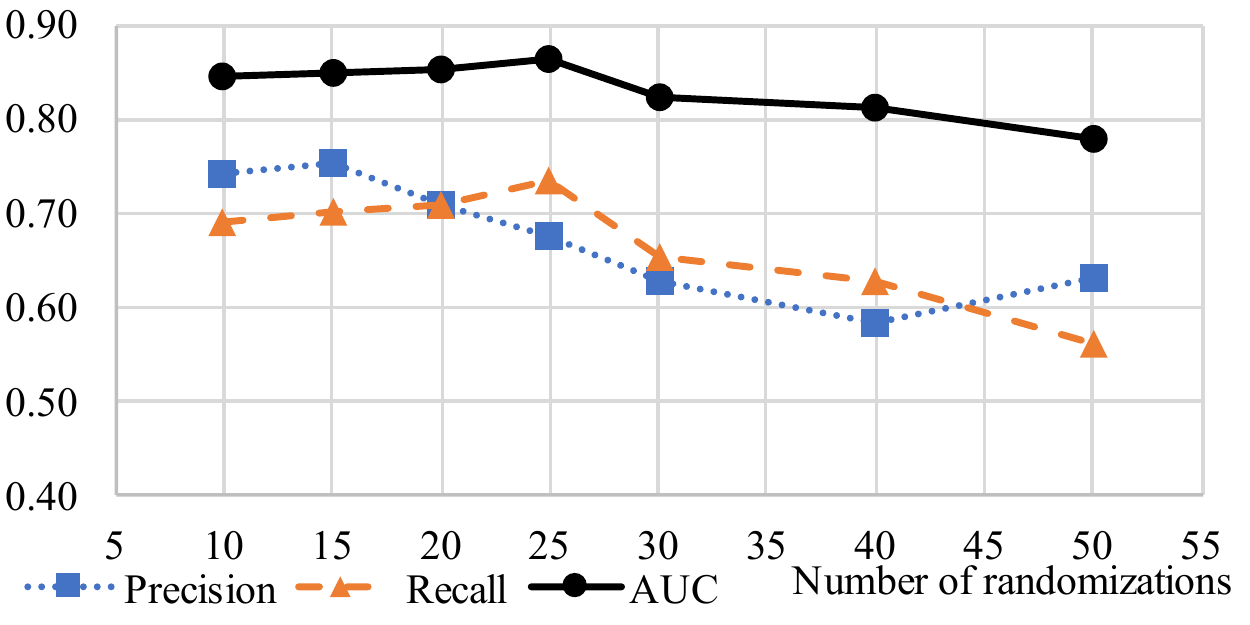}
    \caption{Prediction performance vs. averaged number of accounts per user.\label{fig:randomization}}
\end{figure}

Figure~\ref{fig:randomization} shows how the averaged number of accounts per user affects the prediction performance. In this experiment, we use a subset of the U.S. 2016 dataset with 100 users and 166831 activities, and randomly assign the activities of a user to a varying number of split accounts, $s$, from 10 to 50. AUC is mostly stable when $s$ is between 10 to 25, and starts to drop when $s$ is greater than 25. We also observed that recall drops when $s$ is greater than 25 as well. This is because when the number of accounts per user goes up, the activities from a single user are distributed into too many split accounts, each lacking the characteristics of the user, and thus leads to both inaccuracy in graph embedding and clustering. For the graph  embedding, if two accounts belong to different users, this will make their embedding vectors very similar, and introduce high false positive rate in the label spreading model, thus decrease the precision. For clustering, putting accounts belonged to the same user in different groups, yields high false negative rates, and subsequently low recalls.   \CH{yes this is weak}. 

\subsubsection{Katz Similarity Threshold, $\alpha$}
\begin{figure}
    \centering
    \includegraphics[width=0.47\textwidth]{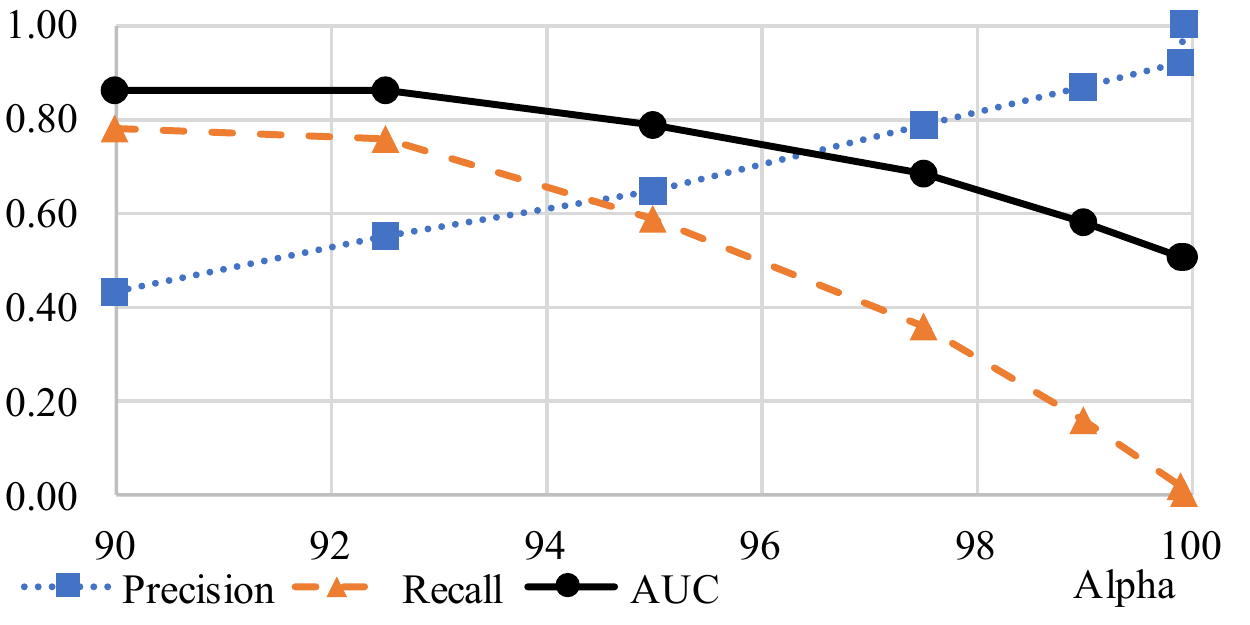}
    \caption{Prediction performance vs. Katz similarity threshold $\alpha$. \label{fig:katz}}
\end{figure}

Figure~\ref{fig:katz} shows how the prediction performance changes with respect to different values of $\alpha$, using Dataset 2 from the Middle East. When increasing $\alpha$ from 90\% to 99.95\%, AUC and the recall decreases, while the precision increases to nearly 1. The empirical threshold for this particular dataset is 95\%, which results in the best combination of AUC, precision and recall. It seems that a few percentiles away from the empirical threshold will not significantly reduce the prediction performance. 

It's clear that if $\alpha$ is set at 99.95\%, all account pairs predicted to be from the same user are indeed from the same user, because the precision is 1. Similarly, if $\alpha$ is set at 80\%, all pairs predicted to be from different users are indeed from different users; our experiment shows that, when $\alpha$ is 80\%, the precision of predicting 0 is 0.997.
As an optimization, we can use predictions with such high confidence as alternative ground truth. Using alternative ground truth can significantly reduce the required number of ground truth queries, and thus reduce the time of getting such ground truth.

\subsubsection{Embedding Vector Length, $d$}
\begin{figure}
    \centering
    \includegraphics[width=0.47\textwidth]{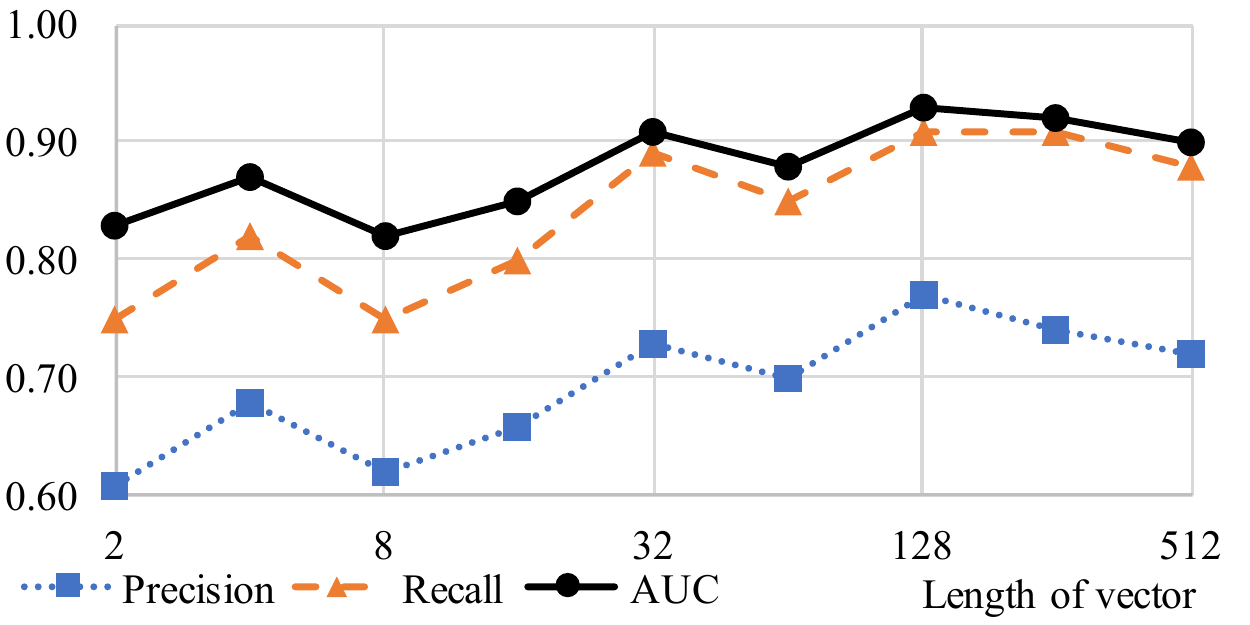}
    \caption{ performance in precision recall and auc in different length of embedding vector \label{fig:diff_d}}
\end{figure}

Figure~\ref{fig:diff_d} shows how the prediction performance changes with respect to the embedding vector length $d$, using Dataset 1 from the Middle East. Generally, increasing the vector length can improve performance, up to some point. When the vector is too short, for instance $d$ less than 32 in this particular experiment, the prediction results are not so good, because the amount of data given to the model is not sufficient. On the other hand, when too much information is present, i.e. $d$ larger than 128, the embedding vector will be too long such that it contains few information in each element, and the prediction performance decreases. The best choice of $d$ is around 128, which is the same as the default value used by Grover et al. ~\shortcite{Grover:2016:NSF}.

\subsubsection{Node2Vec Parameters, $p$ and $q$}
\begin{figure}
    \centering
    \includegraphics[width=0.47\textwidth]{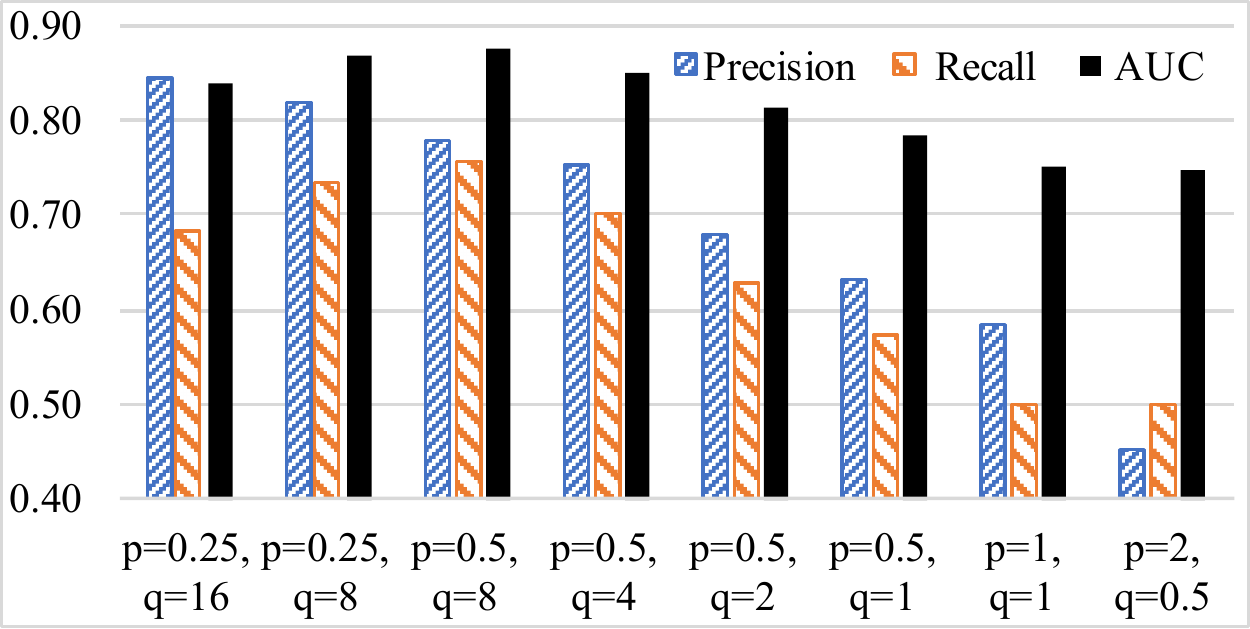}
    \caption{Prediction performance with different $p$ and $q$ values\label{fig:pq}}
\end{figure}
Figure~\ref{fig:pq} shows how the selection of local vs. global information in the embedding affects the prediction performance, using the U.S. 2016 dataset. The graph embedding method, Node2Vec ~\cite{Grover:2016:NSF}, is based on random walks, and it has a pair of parameters $p$ and $q$, with their explanation in the Semi-Supervised Learning using Graph Embedding subsection. 
It is shown that generally, more local information, i.e. lower $p/q$ value (with $p<1, q>1$), yields better performance, which indicates local features are more important than global features in the multiple accounts detection. But there is a threshold, at where $p/q$ is equal to $1/16$, that produces the best AUC and recall. The reason is that, although not as important as local information, global information is also useful; totally ignoring global information will adversely affect the prediction performance.

\subsubsection{Number of Clusters, $c$}
\begin{figure}
    \centering
    \includegraphics[width=0.47\textwidth]{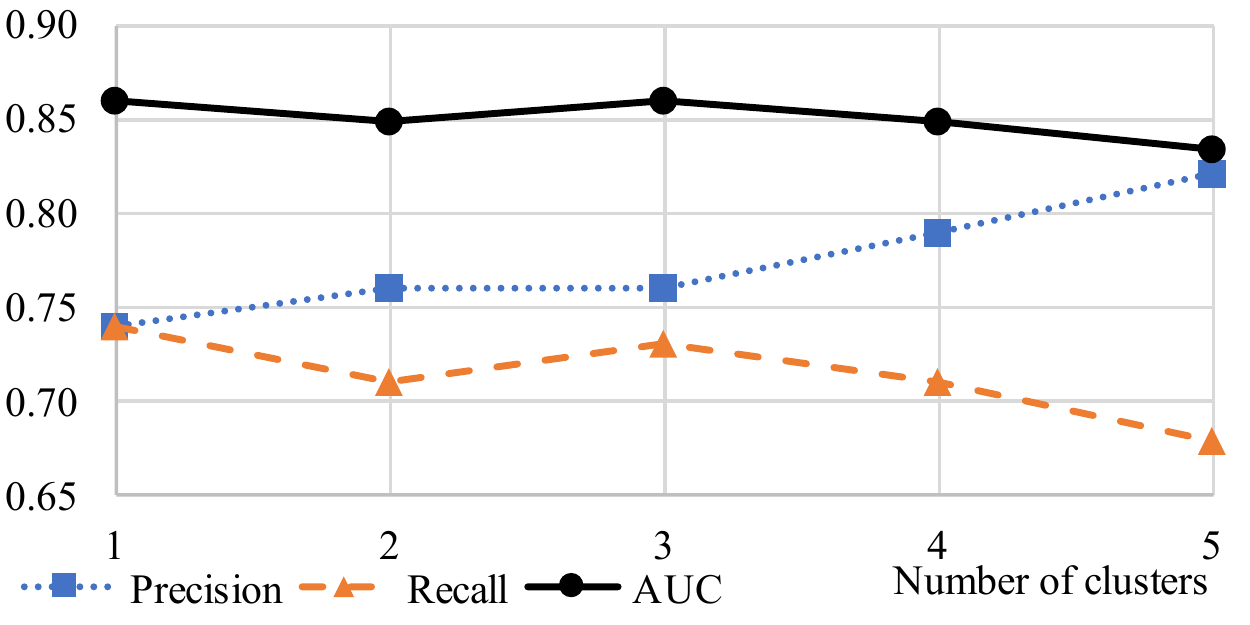}
    \caption{Prediction performance vs. the number of clusters\label{fig:diff_g}}
\end{figure}
Figure~\ref{fig:diff_g} shows how the prediction performance changes with respect to the number of clusters used in the clustering optimization, using Dataset 2 from the Middle East. The number of clusters has light influence on the prediction performance, and no significant change happens when the number of clusters increases from 1. This shows the clustering optimization will not have high impact on the prediction performance, when the number of clusters is small. However, it can significantly reduce the computation workload and the memory requirement of the label spreading algorithm, so as leading to better scalability. However, when there are too many clusters, the recall will decrease, because accounts from the same users are assigned to different clusters.


\subsubsection{With or Without Alternative Ground Truth}
We also test how alternative ground truth will influence the result, using Dataset 2 from the Middle East. We substitute the sampled ground truth with alternative ground truth from Katz similarity (discussed in the Scaling Up the Proposed Algorithms subsection), and validate the predicted results against the actual ground truth. We use three clusters in this experiment, and the precision, recall and AUC are 0.52, 0.76, and 0.86 respectively, compared to 0.74, 0.74 and 0.86 when using the sampled ground truth. Recall and AUC mostly stay the same, but the precision decreases. When getting alternative ground truth for account pairs that should be labelled as coming from the same user, we set $\alpha$ to a very high value, which introduces unbalanced sampling, thus inaccurate cuts, leading to lower precision, and sightly higher recall. This shows using alternative ground truth will not significantly decrease the prediction performance, but can greatly reduce the number of ground truth queries, thus better scalability.

\section{Conclusion and Future Work}
\label{sec:conclusion}
We introduced unsupervised method using Katz similarity, and semi-supervised approaches using Katz similarity and graph embedding, for multiple accounts detection in online social networks. Rather than making predictions to individual accounts, our methods make prediction to each account pair and do not require the exact number of actual users. We also purposed using clustering, and alternative ground truth, to enhance the scalabilty and lower the sample rate of data. Large scale experiments show that our approach works well for multiple accounts detection in different situations. We also explore how graph features and algorithmic parameters, including random walk strategies, will affect the results.

It will worth incorporating graph convolutional network (GCN) features into our method, and constructing graph features in an end-to-end manner. The basic idea is to use an auto-encoder to encode each node, and a decoder to predict whether two nodes in a graph belong to the same user. 
It will also be interesting to find out, whether utilizing other information from the social network, such as time data, location and textual information, can improve the multiple account prediction performance. Similar methods may also be useful to detect malicious accounts. \CH{Too many future work. Remove if we exceed page limit}

\bibliographystyle{aaai}
\bibliography{references}
\end{document}